# Design and Predict Tetragonal van der Waals Layered Quantum Materials of MPd$_5$I$_2$ (M=Ga, In and 3$d$ Transition Metals)

Niraj K. Nepal[1], Tyler J. Slade[1], Joanna M. Blawat[2], Andrew Eaton[3], Johanna C. Palmstrom[2], Benjamin G. Ueland[1,3], Adam Kaminski[1,3], Robert J. McQueeney[1,3], Ross D. McDonald[2], Paul C. Canfield[1,3#] and Lin-Lin Wang[1,3*]

[1]Ames National Laboratory, U.S. Department of Energy, Ames, IA 50011, USA
[2]National High Magnetic Field Laboratory, Los Alamos National Laboratory, Los Alamos, NM 87545, USA
[3]Department of Physics and Astronomy, Iowa State University, Ames, IA 50011, USA

#canfield@ameslab.gov
*llw@ameslab.gov

## Abstract

Quantum materials with stacked van der Waals (vdW) layers hosting non-trivial band structure topology and magnetism have shown many interesting properties. Using high throughput density functional theory calculations, we design and predict tetragonal vdW-layered quantum materials in the MPd$_5$I$_2$ structure (M=Ga, In and 3$d$ transition metals). We show that besides the known AlPd$_5$I$_2$, the -MPd$_5$- structural motif of three-atomic-layer slabs separated by two I layers can accommodate a variety of metal atoms giving arise to topologically non-trivial features and highly tunable magnetic properties in both bulk and single layer 2D structures. Among them, TiPd$_5$I$_2$ and InPd$_5$I$_2$ host a pair of Dirac points and likely an additional strong topological insulator state for the band manifolds just above and below the top valence band, respectively, with their single layers hosting or near quantum spin Hall states. CrPd$_5$I$_2$ is a ferromagnet with a large out-of-plane magneto-anisotropy energy, desirable for rare-earth-free permanent magnets.

# I. Introduction

Design and discovery of novel 2D van der Waals (vdW) layered topological and magnetic materials containing 3$d$ transition metals (TM) are essential for advancing both fundamental science and applied technologies. For example, understanding and manipulating itinerant 3$d$ TM magnetism in intermetallics are crucial to understand many emergent states including superconductivity[1-5]. At the same time, rare-earth-free permanent magnets with 3$d$ TM having large coercivity and saturation moment are sought after for renewable energy technologies[6-11]. When combining these two features in the same material, it is rare to find families of magnetic 2D vdW materials with only a handful examples such as $CrX_3$ (X=I, Br, Cl)[12-17], $VI_3$[18], $Cr_2Si_2Te_6$[19], $Cr_2Ge_2Te_6$[20, 21], $Fe_3GeTe_2$[22] and $MnBi_{2n}Te_{3n+1}$(MBT)[23]. The MBT systems are particularly interesting because most of them are intrinsic antiferromagnetic (AF) topological insulators (TI) that can be exfoliated to a few layers. The discovery of these magnetic 2D vdW materials have galvanized extensive research activities to study their unique properties.

The recent discovery[24-29] of magnetic materials with the structural motif of -$MPd_5$- and -$MPt_5$- slabs in the tetragonal anti-$CeCo_5In$ structure has attracted wide attention for incorporating 3$d$ TM magnetism of Cr, Mn and Fe into slabs of Pd and Pt with large spin-orbit coupling (SOC). Ferromagnetic (FM) $CrPt_5P$[28] and $MnPd_5P$[29] have shown large magneto-anisotropy energy (MAE) for possible rare-earth-free permanent magnet applications, however, the easy axis is in-plane. These 1-5-1 compounds with separation of three-atomic-layered -$MPd_5$- and -$MPt_5$- slabs by a single layer of P or As are not exfoliable due to the strong bonding of P or As to Pd or Pt on both sides. But it hints at a new way to design 2D vdW materials by inserting more anion layers to separate the well-structured atomic metal slabs. On the other hand, non-magnetic (NM) 2D vdW materials that can be exfoliated to a single layer are in high demand to realize a range of emergent quantum states from superconductivity[30] to fractional quantum anomalous Hall effect[31-33] by twisting the few-layer systems, as well as charge density wave (CDW) induced quantum spin Hall (QSH) effect[34] by gating. These new tetragonal vdW-layered materials predicted here can potentially provide a platform for the twisted few-layer systems with a tetragonal lattice[35] besides the twisted hexagonal and orthorhombic lattices.

Here we report the design and prediction of 2D vdW-layered I-separated -MPd$_5$-slabs in the body-centered tetragonal crystal structure of MPd$_5$I$_2$ with (M= Ga, In and 3d TMs) by first-principles calculations. The phase stability calculations based on density functional theory[36, 37] (DFT) show that the NM ones are on the ground state (GS) hull and thermodynamically stable, similar to the already existing AlPd$_5$I$_2$[38], and a few magnetic ones are close to the GS hull and thus metastable. These materials have either topologically non-trivial band structures or interesting magnetic properties. In the two band manifolds near the Fermi energy ($E_F$) separated by the top valence band from above and below, respectively, the NM compounds host a pair of bulk Dirac points (BDPs) just above and likely an additional strong TI state below the top valence band. While TiPd$_5$I$_2$ is a Dirac semimetal (DSM) with a relatively clean Fermi surface (FS), InPd$_5$I$_2$ has a half-filled top valence band with the surface Dirac point (SDP) from the strong TI state appearing at the $E_F$. This combined feature results in the surface states with both the BDP projection and SDP appearing at the Γ point in a narrow energy window of 0.1-0.2 eV around the $E_F$ in TiPd$_5$I$_2$ and InPd$_5$I$_2$. The single layer (1L) TiPd$_5$I$_2$ is a QSH insulator with an indirect global band gap of 30 meV or a narrow-gap semiconductor with a small direct gap of 100 meV depending on the DFT exchange-correlation (XC) functional. The 1L InPd$_5$I$_2$ hosts two QSH states just above and below the half-filled top valence band with the band inversions involving pieces of flat bands. Among the magnetic ones, CrPd$_5$I$_2$ stands out for having the highest MAE of 2.88 meV/f.u. with the FM easy axis along the *c*-axis giving the desirable out-of-plane magnetic anisotropy needed for rare-earth-free permanent magnet. The large MAE remains for the 1L FM CrPd$_5$I$_2$. From the many band crossings around $E_F$ in both bulk and 1L CrPd$_5$I$_2$, there are Weyl nodal lines protected by the horizontal mirror plane as the FM magnetic moment is along the *c*-axis.

## II. Methods

DFT[36, 37] calculations have been performed with different XC functionals using a plane-wave basis set and projector augmented wave method[39], as implemented in the Vienna Ab-initio Simulation Package[40, 41] (VASP). Besides PBEsol[42], for vdW interaction we have used vdW density functional (vdW-DF) of optB86b[43] and the most recent r2SCAN+rVV10[44]. Band structures have been calculated with SOC and the results have also been checked with modified Becke-Johnson[45] (mBJ) and HSE06[46] exchange functional. We have used a kinetic energy cutoff of 400 eV, $\Gamma$-centered Monkhorst-Pack[47] with a k-point density of 0.025 1/Å and a Gaussian smearing of 0.05 eV. The ionic positions and unit cell vectors are fully relaxed with the remaining absolute force on each atom being less than $1\times10^{-2}$ eV/Å. For the 1L structures, ionic relaxation is allowed in all the directions, while the lattice vectors are only relaxed along the in-plane directions (*x-y*) with a 20 Å vacuum inserted along the out-of-plane (*z*) direction. In magnetic systems, calculations are initialized with a magnetic moment of 5 $\mu_B$ for transition metals and 0 $\mu_B$ for other elements in both FM and AF configurations. MAE calculations are performed by changing the global spin quantization axis from the *z* to *x* direction. Phase stability analysis is conducted using the convex hull algorithm[48, 49], which is implemented in the Pymatgen package[50, 51]. The high throughput calculations on electronic structure and thermodynamics have been carried out in the workflow of High throughput Electronic Structure Pakage (HTESP)[52]. To calculate Wilson loop and surface spectral functions, maximally localized Wannier functions (MLWF)[53, 54] and the tight-binding model have been constructed to reproduce closely the band structure within $E_F\pm1eV$ by using Group III *sp*, TM *sd* and I *p* orbitals. The surface spectral functions have been calculated with the surface Green's function methods[55, 56] as implemented in WannierTools[57]. The phonon band dispersions are calculated with the small displacement method as implemented in Phonopy[58].

## III. Results

### III-a. Structural motif and phase stability

Figure 1(a) summaries the design of layered compounds based on the -MPd$_5$- slabs. Without layer separation, the -MPd$_5$- slabs with shared plane boundary is in the Cu$_3$Au-type structure in space group (SG) *Pm-3m* (221) with M being surrounded by 12 nearest-neighbor Pd in a locally close-packed face-centered cubic (FCC) structure. As shown by the successful synthesis of MPd$_5$P and MPd$_5$As compounds[27, 29], an anionic layer can be inserted between the -MPd$_5$- slabs to make new structures in the SG *P4/mmm* (123). With P/As in the nominal valence of 3–, it is possible to have two I to replace P and more importantly it becomes vdW layered, as shown by the existence of AlPd$_5$I$_2$[38], the only reported compound in this structure so far. It replaces P with two I and shifts the -MPd$_5$- slabs in-plane for every other slab, which results in a body-centered tetragonal (tI16) structure in SG *I4/mmm* (139) with a large distance of 3.89 Å between the I atoms of neighboring slabs. A very recent experimental study[59] on single crystal AlPd$_5$I$_2$ has shown that indeed it can be exfoliated to a single AlPd$_5$I$_2$ slab with interesting properties. Here we design and predict new vdW layered compounds in this structure by replacing Al with other group III elements and also 3*d* TMs via high throughput DFT calculations.

As plotted in Fig.1(b) for the projected density of states (PDOS) of NM InPd$_5$I$_2$, NM TiPd$_5$I$_2$ and FM CrPd$_5$I$_2$, most of the I 5*p* orbitals hybridize with the bottom of Pd 4*d* orbitals from –6 to –4 eV to form the *p-d* bonding states. The *p-d* anti-bonding states are pushed to above 1 eV as empty states to gain more cohesion for the ternary compounds with additional bonding hybridization between Pd 5*s* (not shown) and I 5*p* orbitals in the same low-energy range. In contrast to I 5*p*, the *p* orbitals of group III elements at M site, for example In, mostly hybridizes with Pd 4*d* states at a higher energy range (–4 to –2 eV), reflecting their electron positive character. But the states near the $E_F$ in InPd$_5$I$_2$ are dominated by Pd 4*d* and I 5*p* orbitals. Next moving to 3*d* TM, Ti 3*d* orbitals hybridize extensively with Pd 4*d* orbitals giving the broader and lower Pd 4*d*-derived bands than those in InPd$_5$I$_2$ below the $E_F$. There is also a large empty anti-bonding DOS peak just above the $E_F$ due to the 3*d*-4*d* hybridization. Then for Cr with two more 3*d* electrons, these empty states get partially filled and induce a large exchange interaction as shown by the two splitting DOS peaks at 0 and +2 eV. This interaction gives a sizable magnetic moment of 2.80 $\mu_B$ on Cr to prefer FM and importantly the easy axis is along the *c*-axis with a large MAE of 2.88 meV/f.u. The PDOS for other magnetic 3*d* TM MPd$_5$I$_2$ are similar in terms of band hybridizations. The NM TiPd$_5$I$_2$ is an interesting case with the right number of valence electrons that the anti-bonding states are almost completely empty giving a minimum DOS at $E_F$ to form a semimetal, whose topological band structure will be detailed later.

The GS convex hull energy ($E_h$) for all the MPd$_5$I$_2$ compounds studied are plotted in Fig.1(c) for PBEsol[42] and also with vdW exchange functional of optB86b[43]. To confirm that the introduction of I prefers two anion layers instead of one, besides the MPd$_5$I$_2$ in the *I4/mmm* structure, we have also calculated the hypothetical MPd$_5$I in the *P4/mmm* structure. As shown in Fig.1(c), $E_h$ for MPd$_5$I are all above 0.10 eV/atom, much higher than MPd$_5$I$_2$, confirming the qualitative argument that the broken metallic interactions in separating -MPd$_5$- slabs need to be compensated by a strong ionic interaction with enough anionic valence. For the group III elements, Al, Ga and In, the E$_h$ of MPd$_5$I$_2$ are all zero for PBEsol and slightly below 0.01 eV/atom for optB86b, showing they are all thermodynamically stable, because AlPd$_5$I$_2$ has already been found in experiment[38], although optB86b gives a small positive E$_h$ of 0.006 eV/atom. InPd$_5$I$_2$ has the smallest E$_h$, thus the most stable among the group III compounds. For the 3$d$ TMs, first TiPd$_5$I$_2$ is quite stable on the GS hull even for optB86b and we found it is a DSM with a clean FS. Next for V and Cr, E$_h$ becomes positive, but smaller than 0.10 eV/atom. Then E$_h$ decreases for Mn and Fe at the middle of 3$d$ TM series, and increases again for the late 3$d$ Co and Ni, which are the least stable among the 3$d$ TM MPd$_5$I$_2$. Overall, optB86b gives a higher E$_h$ than PBEsol, showing a systematic shift between the different XC functionals. But the trends for the variation in E$_h$ across the whole series for both MPd$_5$I$_2$ and MPd$_5$I are the same for different XC functionals showing the results are well converged.

The magnetic properties across the 3$d$ TM series for MPd5I2 are quite interesting and tabulated at the top of Fig.1(c). Except for Ti being NM, the magnetic moment size increases first starting with V, reaching the maximum of 3.92 $\mu_B$ for Mn before decreases at the end of the series for Co and Ni. With the gradual filling of the 3$d$ orbitals, V, Cr and Mn prefer FM, while Fe, Co and Ni prefer AF. Importantly, both V and Cr prefer easy axis along the $c$-axis with Cr having the largest MAE of 2.88 meV/f.u., much higher than the 0.30 meV/f.u. for VPd$_5$I$_2$ and the rest. In contrast, FM Mn prefers the in-plane easy-axis, although with the largest moment. Then for AF, first Fe moments prefers in-plane and then Co and Ni prefer out-of-plane directions.

To study phase stability and construct GS hull, all the existing binary and ternary compounds together with the elemental ones in the ternary phase diagrams have been computed and their stability are calculated via different possible reaction paths. Although PBEsol gives AlPd$_5$I$_2$ on the GS hull agreeing with experiment in Fig.1(c), Fig.2(a) shows that the PBEsol-calculated volume per atom is underestimated when compared to the available experimental data. Then as shown in Fig.2(b), this underestimation of volume can be improved by using optB86b exchange functional. Also, to explicitly include vdW interaction for the 1-5-2 compounds, we have chosen optB86b vdW exchange functional for the phase stability plots in Fig.2, as well as the band structure and magnetic property calculations. The GS hulls with PBEsol are similar and can be found in Supplementary Figure 1.

As shown in Fig.2, for Pd-I binary, there is only one stable line compound of $PdI_2$. For M-I binaries, there are many stable line compounds for Ga, In, Ti and V. For the rest, there is only one stable binary M-I compound including $CrI_3$ for Cr-I. For M-Pd binaries, Al, Ga, In, Ti, V and Mn have many stable line compounds. Interestingly, for the -$MPd_5$- motif in $MPd_3$ or the $Cu_3Au$-type, this binary line compound structure exists for In, Ti, V, Cr, Mn and Fe with Pd. But for Ni and Co, they form random alloy or solid solution with Pd. Because both $CoPd_3$ and $NiPd_3$ are only slightly above the GS hull at 0.06 eV/atom, they can be used as good representatives for the binary solid solution and we include them in the phase stability calculation of $MPd_5I_2$. Among the calculated $MPd_5I_2$, the NM compounds are more stable than the magnetic ones. Given $AlPd_5I_2$ with the $E_h$ of 0.006 eV/atom in optB86b has already been synthesized, we predict the existence of $GaPd_5I_2$, $InPd_5I_2$ and $TiPd_5I_2$ because they are on the GS hull. For magnetic ones, Mn and Fe have the smallest $E_h$ above GS hull and then followed by V and Cr. Considering the approximations used in DFT calculations, we predict these four magnetic ternaries are metastable, also because the binary $MPd_3$ line compounds with the -$MPd_5$- motif are stable and found in experiments. Lastly for Co and Ni, they have the largest $E_h$ above GS hull even with PBEsol and due to the solid solution of $MPd_3$, it is also possible to form solid solution for the ternary compounds. We predict these two are possible ternaries but with solid solution tendency, which needs to be further studied in the future.

From an experimental viewpoint, these $MPd_5I_2$ compounds are much more challenging to synthesize than $MPd_5P$ and $MPd_5As$ because of the higher vapor pressure or lower sublimation temperature of I than P and As. We have also studied the dynamical stability of these predicted compounds by calculating the phonon band dispersion. For example, the phonon band dispersions of both bulk and 1L structures for $TiPd_5I_2$, $InPd_5I_2$ and $CrPd_5I_2$ are plotted in Supplementary Figure 2. The absence of imaginary phonon modes shows that these predicted compounds in bulk and 1L structures are dynamically stable.

## III-b. Topological features of non-magnetic 1-5-2 compounds

For the band structures of NM MPd$_5$I$_2$, we chose TiPd$_5$I$_2$ and InPd$_5$I$_2$ to present the topological band features of both the bulk and 1L structures. Bulk band structures of other MPd$_5$I$_2$ can be found in Supplementary Figure 3. Figure 3(a) plots the band structure of bulk TiPd$_5$I$_2$ without SOC with the body-centered tetragonal Brillouin zone (BZ) and high symmetry k-points shown in Fig.3(c). The highest valence band (N) and band below (N–2) according to simple filling are shown in red and blue, respectively, with N being the number of valence electrons. Above the highest valence band, there is a sizable gap in most of the BZ, except for around the Z point in the Γ-Z, Z-S$_1$ and Y$_1$-Z directions. Along the Γ-Z direction, band N and N–2 are degenerate, giving triple degeneracy (or six-fold including spin) at the crossing point between the highest valence and lowest conduction band in middle of the Γ-Z direction. From the triple degeneracy point to the Z point, a doubly degenerated nodal line segment appears, as also shown in Fig.3(d) by plotting the zero-gap k-points in the whole BZ. The crossings along the Z-S$_1$ and Y$_1$-Z directions are parts of the nodal line loops around the Z point on the (110) and (1–10) planes, which are protected by the diagonal mirror symmetries.

With SOC, as plotted in Fig.3(b), the orbital degeneracy for the top two valence bands along the Γ-Z direction is lifted and the nodal loops are all gapped out, except for the crossing between the highest valence and lowest conduction band along the Γ-Z, forming a BDP as protected by the four-fold rotational symmetry. Because of the time-reversal symmetry (TRS) and inversion symmetry, each band is still doubly degenerated. The BDP is zoomed in Fig.3(e) along the Γ-Z direction showing the zero gap and the switching between I $p_z$ and Pd $d_{xz}/d_{yz}$ orbitals with the 2-dimensional irreducible representations of Γ$_9$ and Γ$_6$. The BDPs are at the momentum-energy of (0, 0, ±0.1499 Å$^{-1}$; E$_F$+0.0236 eV), as also shown by the red dots in Fig.3(c). The band dispersion of TiPd$_5$I$_2$ is zoomed along the Z-S$_1$ direction in Fig.3(f) to show the small SOC-induced gap.

Additionally, for the highest valence band N (red), it is also gapped from below by the next valence band N–2 (blue). The lower branch of the band N–2 along the Γ-Z direction forms a band inversion region around the Z point with the top valence band N. Wilson loop calculation shows that this N–2 band manifold hosts a strong TI (STI) state with the Fu-Kane[60] topological index of (1;001). The non-trivial Z$_2$ number is shown by the Wilson loop of Wannier charge centers (WCC) on the $k_z=0.5$ plane in Fig.3(g) with the odd number of crossings by the dashed line with WCC. Calculations with the more recent r2SCAN+rVV10[44] XC functional give similar band features (see Supplementary Figure 4) for both DSM and STI in $TiPd_5I_2$. For mBJ[45] functional, the BDP still remains and predicts a DSM, despite most of the valence bands being pushed lower and conduction bands higher in energy. But the extra band inversion at the Z point is lifted between band N–2 and N. The existence of this extra band inversion for STI or not in $TiPd_5I_2$ will be the features need to be verified in experiment.

To demonstrate the non-trivial band structure of $TiPd_5I_2$ with both a BDP above and a STI below the highest valence band, we have calculated the (001) surface spectral functions using the Wannier functions. On (001) surface, the projections of the BDP at $\pm k_z$ onto the same $\bar{\Gamma}$ point stands out nicely around the $E_F$ because of no overlap with other bulk band projection for the clean FS. There are topological surface states (TSS) stemming from the BDP projection as seen in Fig.3(h) at $E_F+0.03$ eV. Below that at $E_F–0.2$ eV is the SDP from the STI of the N–2 band manifold also shown clearly even though on top of the other bulk band projections. The spin-texture of the surface Dirac cone is plotted at $E_F–0.185$ eV in Fig.3(j) confirming the spin-momentum locking of the surface Dirac cone. The spin-momentum locking of the TSS stemming from the BDP projection is shown in Fig.3(i) at $E_F+0.045$ eV. Thus, bulk $TiPd_5I_2$ is a DSM with a clean FS and also possibly hosts a STI just below the highest valence band.

When the vdW-layered $TiPd_5I_2$ is exfoliated down to 1L, the band structure is plotted in Fig.4(a). A large band gap exists in most of the BZ, while a small gap appears along the Γ-X direction, which is projected from the Z-$S_1$ direction from the bulk band structure. Overall there is an indirect global band gap of 30 meV between the valence band maximum at X and conduction band minimum at Γ point. The Wilson loop calculation of the highest valence band manifold (red) in Fig.4(b) indicates a QSH with an odd number of crossings of WCC. In contrast, for the N–2 band manifold (blue), the even number of crossings in the Wilson loop in Fig.4(c) shows it is topologically trivial. The edge spectral functions are plotted in Fig.4(d) and (e) for the different TiPd- and PdI-terminations, respectively. The topological edge states connect the gapped valence with conduction band projections to form a TRS-protected edge Dirac points (EDP) at the $\bar{\Gamma}$ point. While the EDP on the TiPd-termination is inside the QSH gap, that on PdI-termination is merged into the valence band projection at $E_F$–0.06 eV. With $E_F$ cutting through the QSH gap, the spin-momentum locked edge states are unavoidable from TRS topological protection. Calculations with r2SCAN+rVV10 XC functional give similar band inversion feature at the Γ point for QSH (see Supplementary Figure 4). In contrast, HSE06[46] functional pushes the valence band lower and conduction bands higher in energy, and lift the band inversion and changes the indirect band gap to a direct gap of just 100 meV at the Γ point (see Supplementary Figure 4(d)). Such a small gap size can be potentially tuned to close and reopen by strain to induce the band inversion for a topological phase transition to realize a critical 2D Dirac point at the Γ point and then a QSH.

Next for $InPd_5I_2$ as an example from group III $MPd_5I_2$, its bulk band structure with SOC is plotted in Fig.5(a). Due to TRS and inversion symmetry, each band is doubly degenerated. Because of the odd number of electrons, the highest valence band (red in Fig.5(a)) is only half-filled, indicated by $E_F$ sitting right in the middle of the band width. But it is still meaningful to discuss the topological features of the band manifolds above and below this half-filled top valence band with a finite FS. Similarly to $TiPd_5I_2$, between the highest valence and lowest conduction band of $InPd_5I_2$, there is only one crossing along the Γ-Z direction as protected by the 4-fold rotational symmetry for a pair of BDP. The BDP is zoomed in Fig.5(b) shown by the projection on I $p_z$ and Pd $d_{xz}/d_{yz}$ orbitals with the 2-dimensional irreducible representations of $Γ_9$ and $Γ_6$. The BDP is at the momentum-energy of (0, 0, ±0.0406 Å$^{-1}$; $E_F$+0.0955 eV). In contrast, between the highest and the next valence band (blue), there is no band crossing. For such gapped band manifolds, the Fu-Kane topological index has been calculated as (1;001) showing a STI state with the Wilson loop of WCC at the $k_z$=0.5 plane being plotted in Fig.5(c). To confirm the STI with surface spectral function on (001), the surface Dirac cone is shown clearly in Fig.5(d) with the SDP right at the $E_F$ inside the projected bulk gap. The spin-texture of the surface states at the $E_F$ is plotted in Fig.5(e) confirming the spin-momentum locking topological feature without the overlap with bulk band projection. In contrast, the BDP projection at $E_F$+0.10 eV on (001) surface in Fig.5(d) is buried inside the other bulk band projection and shows no TSS, unlike those in $TiPd_5I_2$ with a clean FS. Thus, group III $MPd_5I_2$ host a BDP above and a STI below the half-filled highest valence band. For $InPd_5I_2$, both the SDP and BDP projection appear at the $\bar{\Gamma}$ point on (001), and they are also within an energy window of 0.1 eV with the SDP being right at the $E_F$ and the BDP just above the $E_F$.

The band structure of 1L InPd$_5$I$_2$ is plotted in Fig.5(f). Again, due to the odd number of electrons, the top valence band is half-filled with $E_F$ sitting in the middle of the band width. But the valence band is continuously gapped from both below and above with band inversions, so topological properties can be calculated. The Wilson loop calculations of the band manifolds in Fig.5(g) and (h) show the odd number of crossings of WCC confirming it hosts two QSH states. The edge spectral functions are plotted in Fig.5(i) and (j) for two different terminations, which are rather similar. The TRS-protected EDP at $E_F$+0.35 eV is for the upper QSH and the EDP at $E_F$ is for the lower QSH. The topological edge states of the upper QSH around the $\bar{\Gamma}$ point stand out, which is accessible, when the $E_F$ can usually be tuned by doping and gating. Calculations with r2SCAN+rVV10, mBJ and HSE06 XC functionals all give the similar two band inversions and topological features (see Supplementary Figure 5), which are much less affected than TiPd$_5$I$_2$, because InPd$_5$I$_2$ bands are more metallic from the half-filled top valence band than TiPd$_5$I$_2$. So 1L InPd$_5$I$_2$ hosts two QSH states in a tetragonal structure despite being a metal. Additionally, the band inversions for the two QSH states around the $E_F$ in Fig.5(f) involve pieces of flat bands from the orbital-decorated square lattice as recently proposed[59]. Together with the QSH insulator with a small indirect band gap or a narrow-gap semiconductor in 1L TiPd$_5$I$_2$, the few layer tetragonal systems of these exfoliable 1-5-2 compounds will be an interesting playground for emergent quantum states in future studies.

### III-c. Magneto-anisotropy of CrPd$_5$I$_2$

Among the magnetic 3$d$ TM MPd$_5$I$_2$ compounds, CrPd$_5$I$_2$ has the largest MAE and also the easy-axis is along the $c$-axis. First without SOC, the spin DOS of the bulk and 1L CrPd$_5$I$_2$ are plotted in Fig.6(a) and (b), respectively. While the spin down (majority) forms a pseudo gap near the $E_F$, the spin up (minority) has a local DOS maximum at $E_F$. Via the hybridization between Cr-3$d$ and Pd-4$d$ orbitals to form bonding and anti-bonding states just above $E_F$, the exchange splitting interaction gives a sizable magnetic momentum of 2.8 $\mu_B$ on the Cr atom. The 1L spin DOS in Fig.6(b) is similar and has narrower and sharper peaks than the bulk in Fig.6(a) due to the smaller band dispersion from the reduced interlayer interactions. To analyze the origin of the large MAE in $CrPd_5I_2$, we have calculated the *k*-point-resolved MAE over the entire BZ by fixing the magnetic charge density but rotating the magnetic axis from [001] to [100] with SOC. As shown in Fig.6(c) and (d) for the $\Delta$MAE=±0.03 meV/f.u., respectively, the positive MAE contribution (favoring the *c*-axis) in Fig.6(c) is mostly around the Γ and X points. In contrast, the negative MAE contribution (favoring in-plane) in Fig.6(d) is mostly from the Z, S point and also half way between the Γ and X points. Going to the 1L $CrPd_5I_2$, the whole band width is reduced, but most of the hybridization peaks remain in the same positions, which shows that the 1L can retain the chemical stability. The magnetic moment does not change much, the MAE is still as high as 2.47 meV/f.u.

The band structures of FM bulk and 1L $CrPd_5I_2$ with SOC, are plotted in Fig.6(e) and (f), respectively. The band double degeneracies are all lifted with the top valence band shown in red. There are many bands crossing the $E_F$, indicating a more complicated FS than the NM $TiPd_5I_2$ and $InPd_5I_2$. These many crossings form 2-fold degenerated Weyl nodal lines as plotted in Fig.6(g) and (h). For the FM bulk $CrPd_5I_2$, besides the main Weyl nodal loops on the $k_z=\pm0.5$ plane, there are also loops around the X points. For the FM 1L $CrPd_5I_2$, there are three Weyl nodal loops, one around the X point and two around the M point. These Weyl nodal lines are within $E_F\pm0.2$ eV.

The high MAE in $CrPd_5I_2$ reflects the unique structural motif of the -$MPd_5$- slab, where each moment-bearing 3*d* TM atom is surrounded by Pd with much larger SOC strength. The distance among the 3*d* TM atoms is much larger than that in elemental solids. The magnetic coupling among the 3*d* TM atoms are through Pd with a larger SOC and itself is near the Stoner magnetic instability. Such combination gives a range of magnetic configurations in $MPd_5I_2$. With the gradual filling of the 3*d* orbitals. V, Cr and Mn prefer FM, while Fe, Co and Ni prefer AF. Importantly, both V and Cr prefer easy axis along the *c*-axis with Cr having the largest MAE of 2.88 meV/f.u. In contrast, FM Mn prefers the in-plane easy-axis, although with a larger moment. Then for AF, first Fe prefers in-plane and then Co and Ni prefer out-of-plane. Among the FM ones, with the largest MAE and easy-axis being out-of-plane, $CrPd_5I_2$ can give a large coercivity field, which is attractive for developing rare-earth-free permanent magnets.

# IV. Discussion

Using high throughput density functional theory calculations, we have explored the phase stability, topological and magnetic properties of MPd$_5$I$_2$ compounds (M=Ga, In and 3$d$ TM), a family of -MPd$_5$- slabs separated by two layers of I to design and predict new vdW-layered quantum materials with tetragonal structure. After confirming the existing AlPd$_5$I$_2$ is on the ground state (GS) hull, we find non-magnetic (NM) compounds with M=Ga, In and Ti are also on the GS hull and thermodynamically stable. For the magnetic ones with 3$d$ TM, we find V, Cr, Mn and Fe are not far above the GS hull and metastable, given the existence of the binary structures with -MPd$_5$- slab in the cubic MPd$_3$ structure. For Co and Ni, the hull energy is the largest and also these MPd$_3$ form random alloys. Using TiPd$_5$I$_2$ and InPd$_5$I$_2$ as examples, we show that the NM MPd$_5$I$_2$ host a pair of bulk Dirac points for the band manifolds just above the highest valence band and also likely an extra strong topological insulator (TI) state from below, respectively. While TiPd$_5$I$_2$ is a Dirac semimetal with a mostly clean Fermi surface, InPd$_5$I$_2$ has a half-filled top valence band with the surface Dirac point from the strong TI appearing at the Fermi energy ($E_F$). This combination gives the (001) surface hosting both a surface Dirac point and a bulk Dirac projection just 0.1-0.2 eV separation at the $\bar{\Gamma}$ point. From different exchange-correlation functionals, the 1L TiPd$_5$I$_2$ is either a quantum spin Hall (QSH) insulator with an indirect global band gap of 30 meV or a narrow-gap semiconductor with a direct gap of 100 meV, which potentially can be tuned for a topological phase transition. In contrast, the 1L InPd$_5$I$_2$ is always a metal from half band-filling and hosts two QSH states with pieces of the flat bands near $E_F$. For the magnetic MPd$_5$I$_2$ with 3$d$ TMs, the preferred magnetic ground state changes with the gradual filling of the 3$d$ orbitals. V, Cr and Mn prefer a ferromagnetic (FM) ground state with less or at the half-filling, while Fe, Co and Ni prefer anti-ferromagnetic configuration with more than half-filling of the 3$d$ orbitals. Interestingly both VPd$_5$I$_2$ and CrPd$_5$I$_2$ have their FM moment easy axis along the out-of-plane $c$-axis, a desirable feature to develop rare-earth-free permanent magnets. CrPd$_5$I$_2$ has a large MAE of 2.88 meV/f.u. Thus, our calculations predict that a group of MPd$_5$I$_2$ are synthesizable tetragonal vdW-layered quantum materials with either non-trivial topological features or highly tunable magnetic properties.

**Data Availability**: The data that support the findings of this study are available from the corresponding authors upon reasonable request.

## Acknowledgements

The topological band structure calculations and analysis were supported by the Center for the Advancement of Topological Semimetals, an Energy Frontier Research Center funded by the U.S. Department of Energy Office of Science, Office of Basic Energy Sciences through the Ames National Laboratory under its Contract No. DE-AC02-07CH11358. The phase stability and magneto-anisotropy calculations in this work at Ames National Laboratory were supported by the U.S. Department of Energy, Office of Science, Basic Energy Sciences, Materials Sciences and Engineering Division. The Ames National Laboratory is operated for the U.S. Department of Energy by Iowa State University under Contract No. DE-AC02-07CH11358. Some of the phonon calculations used resources of the National Energy Research Scientific Computing Center (NERSC), a DOE Office of Science User Facility. The National High Magnetic Field Laboratory is supported by National Science Foundation through NSF/DMR-2128556 and the State of Florida.

**Author Contributions**: L.-L.W. and P.C.C. conceived and designed the work with inputs from T.J.S. and N.K.N. N.K.N. and L.-L.W. performed the ab initio calculations on phase stability, topological band structure analysis, magneto-anisotropy and phonon dispersion. J.M.B., A.E., J.C.P., B.G.U., A.K., R.J.M., and R.D.M. contributed to discuss and validate the results. All the authors contributed to the writing and review of the final manuscript.

**Competing Interests**: The authors declare no competing interests.

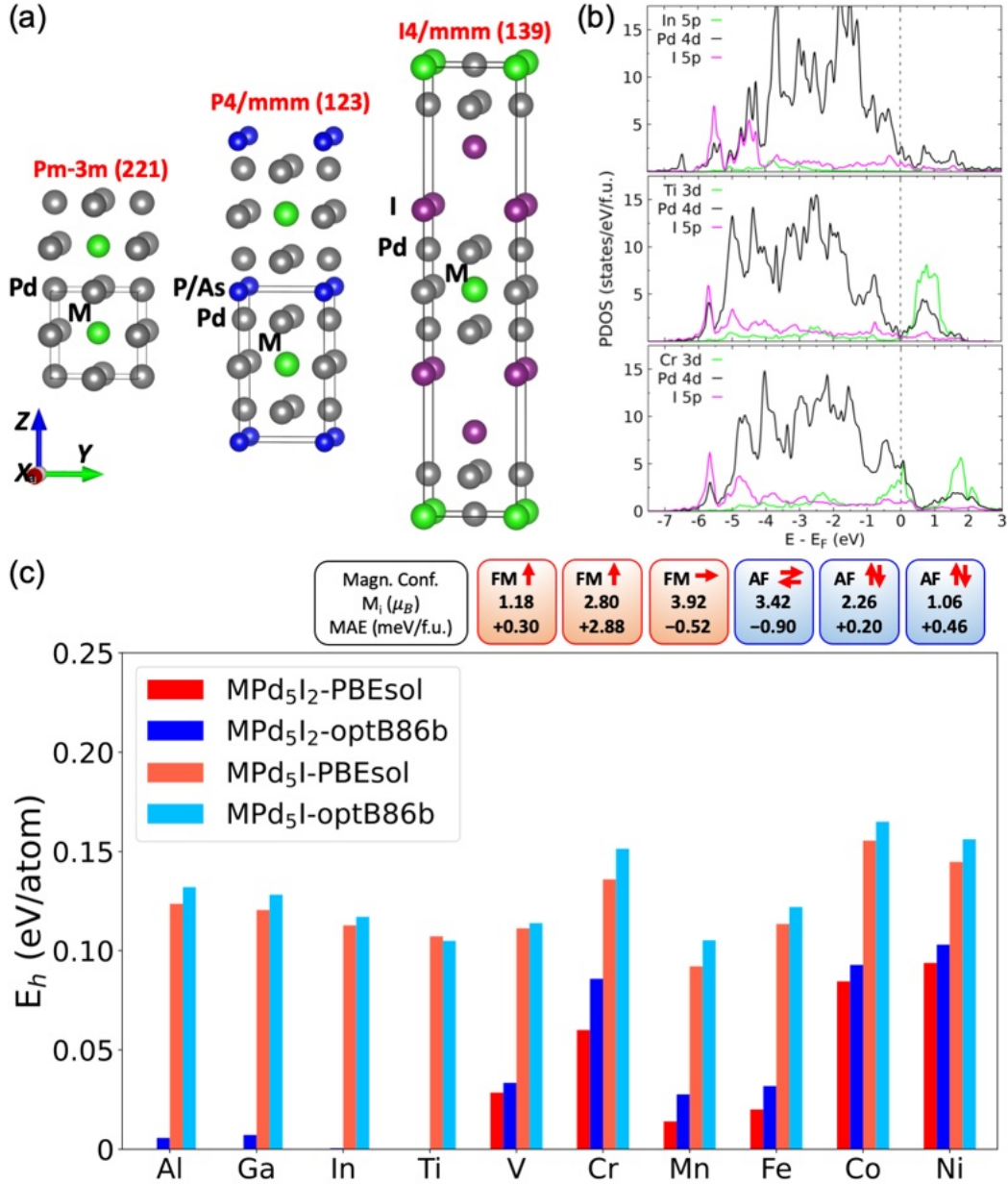

Figure 1. Summary of MPd$_5$I$_2$ structural motifs, electronic and magnetic structures. (a) Crystal structures of MPd$_5$ motif in the three-atomic-layer slab with increasing distance between slabs as in MPd$_3$ of *Pm-3m* (221) in Cu$_3$Au-type, MPd$_5$P of *P4/mmm* (123) and MPd$_5$I$_2$ of *I4/mmm* (139). The atomic species are shown in different colors and labeled accordingly. (b) Projected density of states (PDOS) on atomic orbitals of non-magnetic InPd$_5$I$_2$, TiPd$_5$I$_2$ and ferromagnetic (FM) CrPd$_5$I$_2$. (c) Convex hull energy ($E_h$) of MPd$_5$I$_2$ and MPd$_5$I (M=Al, Ga, In, Ti, V, Cr, Mn, Fe, Co and Ni) calculated in density functional theory (DFT) with PBEsol and optB86b exchange-correlation functionals. The ground state magnetic configurations with easy axis are drawn with the listed moment size on the magnetic ions (from V to Ni) and magneto-anisotropy energy (MAE). FM and anti-ferromagnetic (AF) are shown in red and blue shaded squares, respectively.

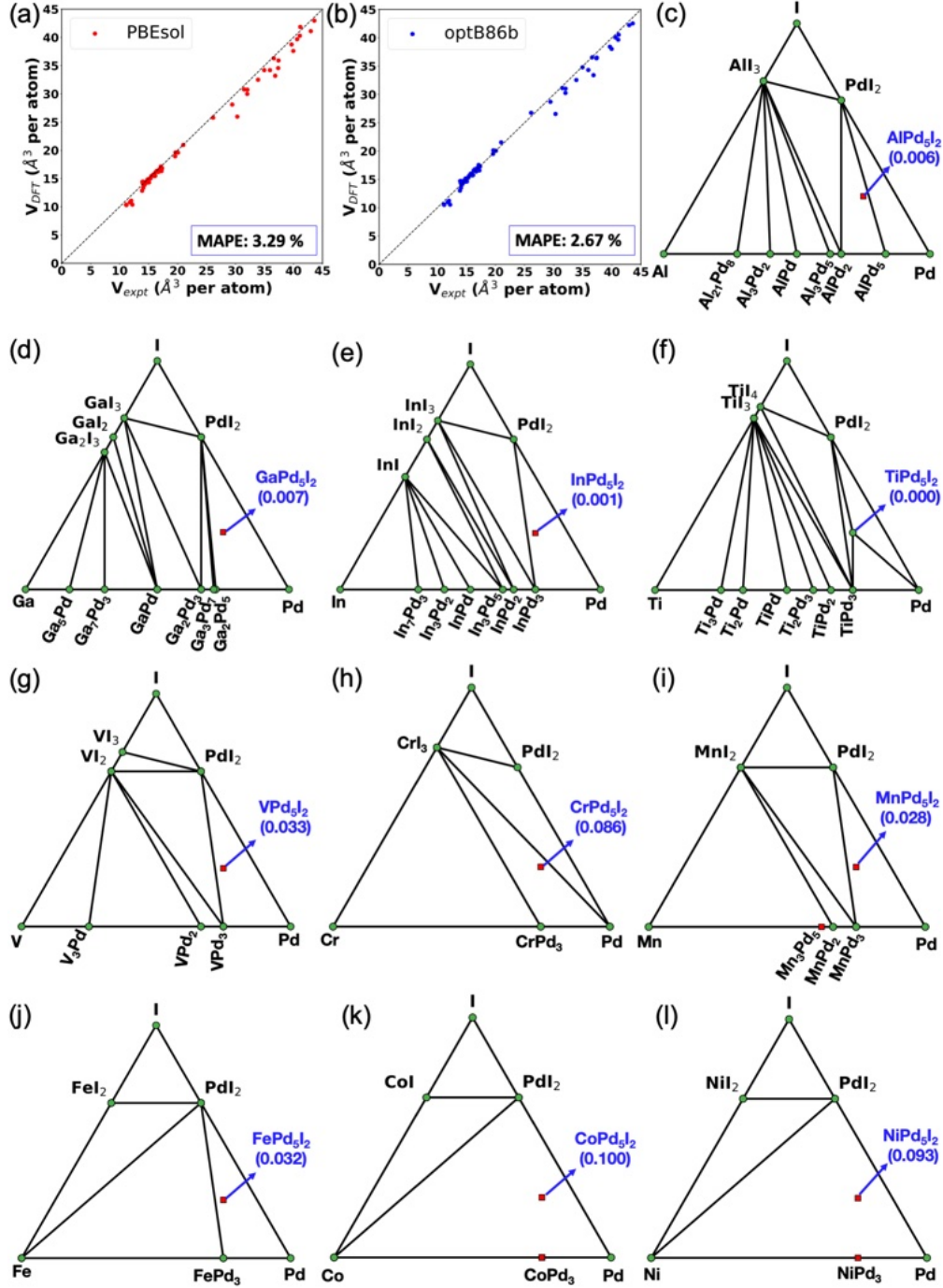

Figure 2. Phase stability and structural energies of $MPd_5I_2$. (a) Volume per atom of the fully relaxed elemental, binary and ternary compounds in PBEsol are compared to available experimental data with the mean absolute percentage errors (MAPE) listed. (b) Same comparison for optB86b. (c)-(l) The calculated phase stability and structural energies of $MPd_5I_2$ (M=Al, Ga, In, Ti, V, Cr, Mn, Fe, Co and Ni) in optB86b. The compounds on the ground state hull are labeled as green dots. The $MPd_5I_2$ are indicated with arrows and their respective hull energies are listed in parenthesis.

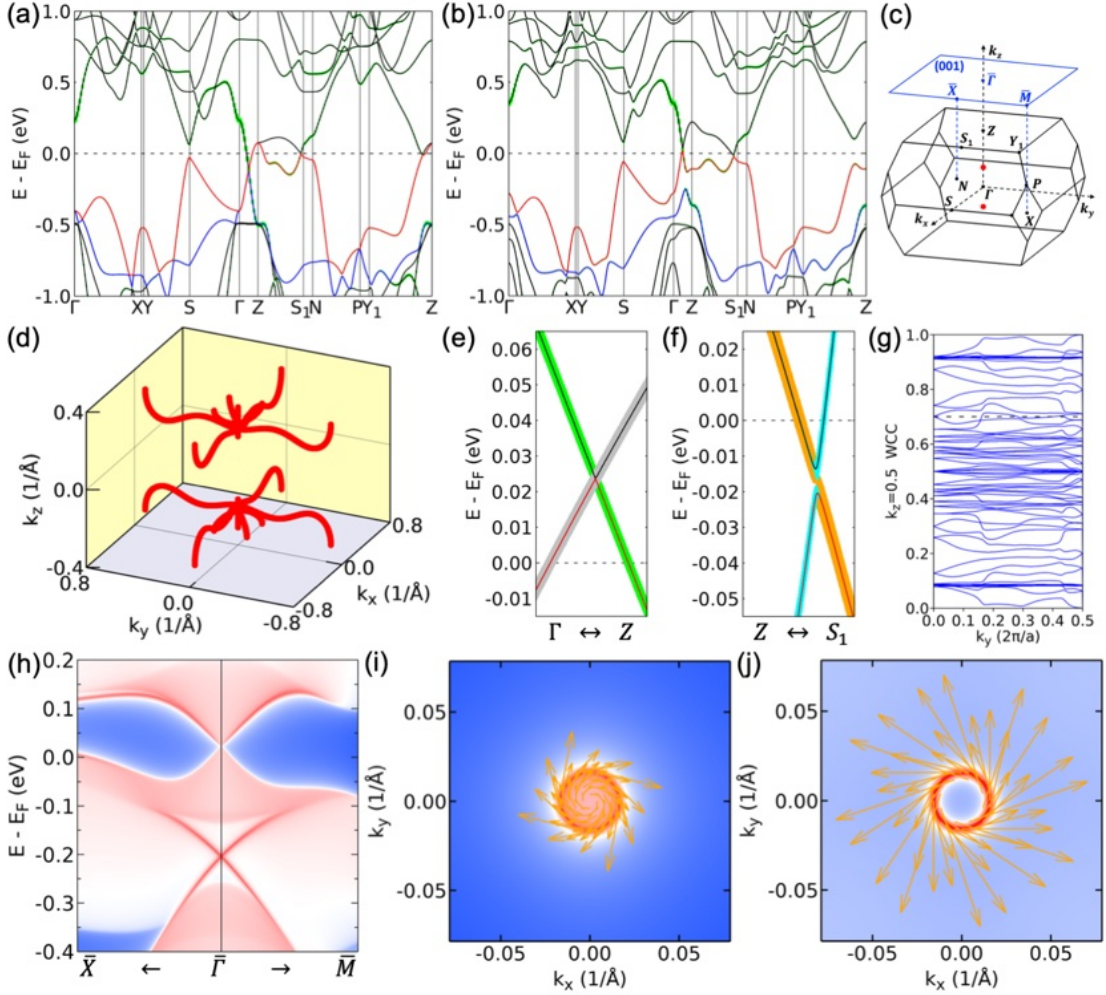

Figure 3. Electronic band structures and topological features of bulk $TiPd_5I_2$. (a) Band structure of $TiPd_5I_2$ without spin-orbit coupling (SOC) with the highest valence band (N) and band below (N-2) shown in red and blue, respectively. Green shade stands for I $p_z$ orbital projection. (b) Band structure of $TiPd_5I_2$ with SOC highlighting the top two valence bands and same orbital projection as in (a). (c) Bulk Brillouin zone (BZ) with high symmetry $k$-points and those on (001) surface BZ are labeled. The bulk Dirac points (BDP) along $\Gamma$-$Z$ are indicated by red dots. (d) Nodal lines in $TiPd_5I_2$ without SOC between the highest valence and lowest conduction bands. (e) Bands zoomed in along $\Gamma$-$Z$ around the BDP between the highest valence and lowest conduction band. The green (grey) shade is for I $p_z$ (Pd $d_{xz}$ and $d_{yz}$) orbital. (f) Bands zoomed in along the $Z$-$S_1$ direction with a small gap opening. The orange (cyan) shade is for Pd $d_{yz}$ (Ti $d_{xz}$) orbital. (g) Wilson loop of Wannier charge centers (WCC) on the $k_z$=0.5 plane between band N–2 and N. (h) Surface spectral function along $\bar{X}$-$\bar{\Gamma}$-$\bar{M}$ on (001). (i) (001) 2D surface Fermi surface at Fermi energy ($E_F$)+0.045 eV and (j) $E_F$-0.185 eV with spin texture shown in orange arrows.

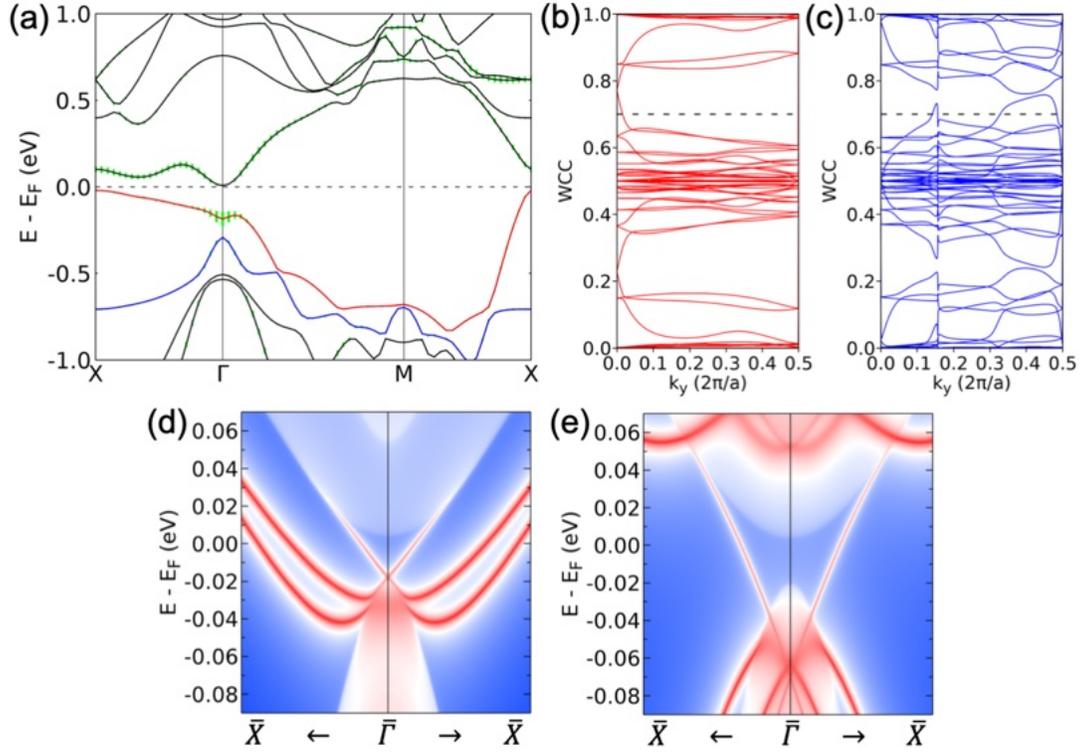

Figure 4. Electronic band structures and topological features of single-layer (1L) TiPd$_5$I$_2$. (a) Band structure of 1L TiPd$_5$I$_2$ with the highest valence band (N) and band below (N-2) shown in red and blue, respectively. (b) Wilson loop of Wannier charge center (WCC) between band N and N+2. (c) Wilson loop between band N–2 and N. (d) Edge spectral functions along $\bar{X}$-$\bar{\Gamma}$-$\bar{X}$ on TiPd- and (e) PdI-terminations.

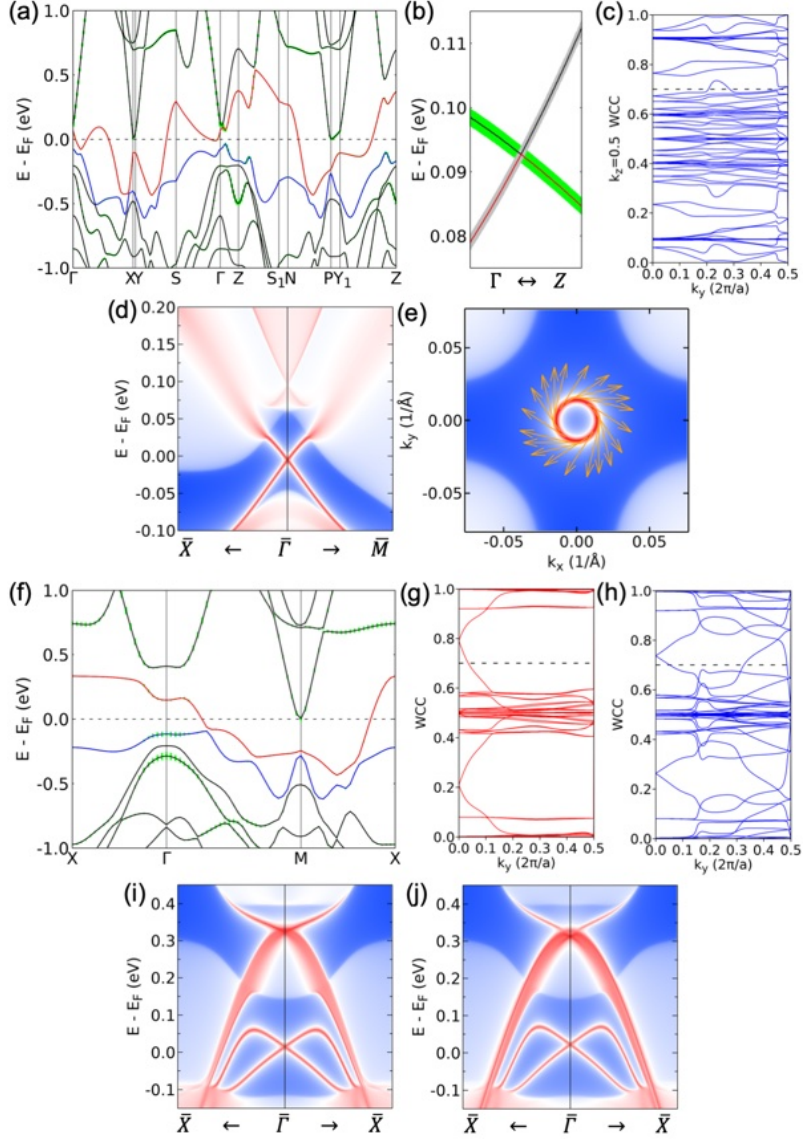

Figure 5. Electronic band structures and topological features of bulk and single-layer (1L) InPd$_5$I$_2$. (a) Band structure of bulk InPd$_5$I$_2$ with the highest valence band (N) and band below (N-2) shown in red and blue, respectively. (b) Bands zoomed in along $\Gamma$-$Z$ around the Dirac point (DP) between the highest valence and lowest conduction band. The green (grey) shade is for I $p_z$ (Pd $d_{xz}$ and $d_{yz}$) orbital. (c) Wilson loop of Wannier charge center (WCC) on the $k_z$=0.5 plane between band N–2 and N. (d) Surface spectral function along $\bar{X}$-$\bar{\Gamma}$-$\bar{M}$ on (001). (e) (001) 2D surface Fermi surface at Fermi energy (E$_F$) with spin texture shown in orange arrows. (f) Band structure of 1L InPd$_5$I$_2$ with the highest valence band (N) and band below (N-2) shown in red and blue, respectively. (g) Wilson loop of WCC between band N and N+2. (h) Wilson loop between band N–2 and N. (i) Edge spectral functions along $\bar{X}$-$\bar{\Gamma}$-$\bar{X}$ on InPd- and (j) PdI-terminations.

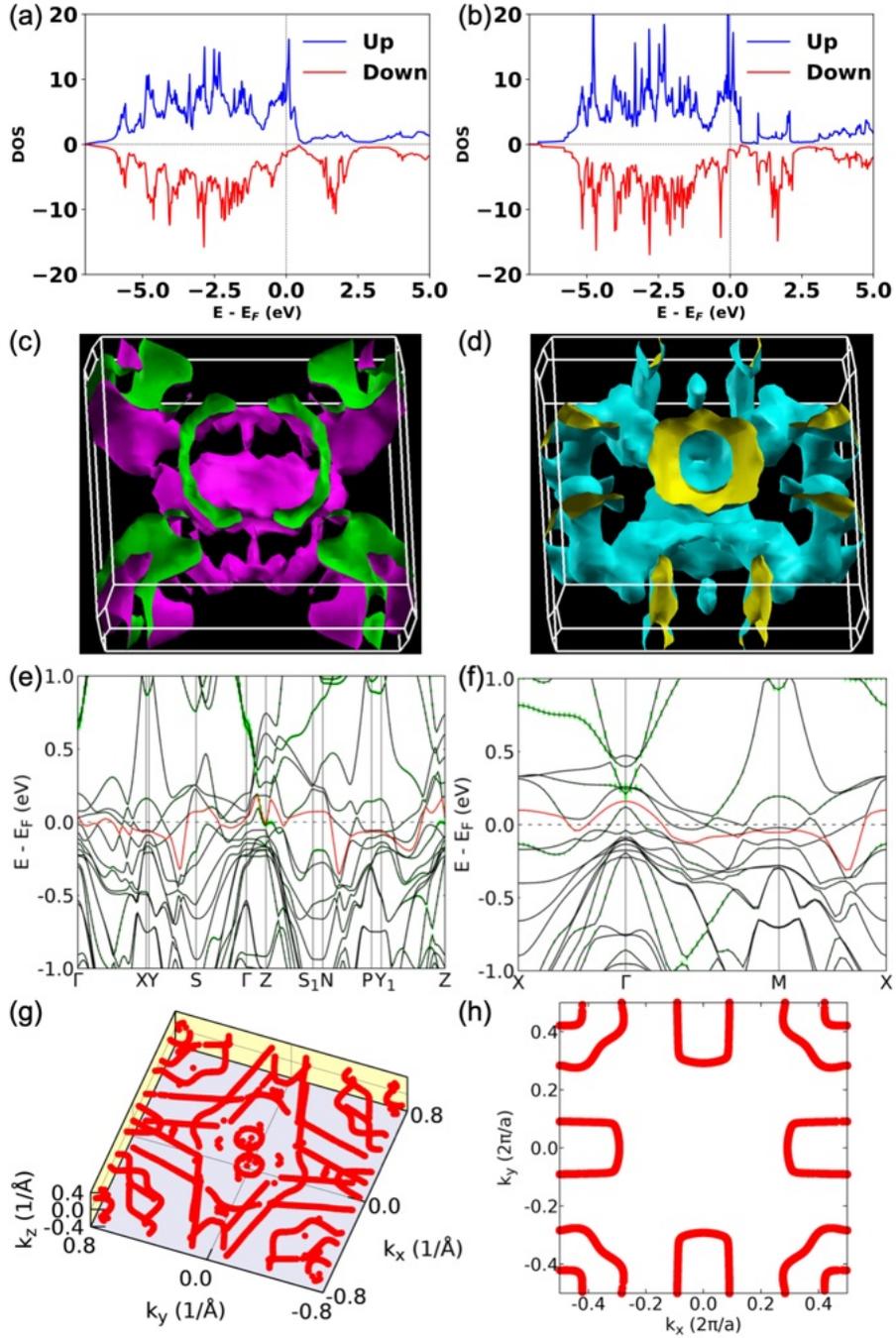

Figure 6. Electronic structures and crystalline magneto-anisotropy of bulk and single-layer (1L) $CrPd_5I_2$. (a-b) Up (minority) and down (majority) density of state (DOS) for bulk and 1L $CrPd_5I_2$ without spin-orbit coupling (SOC) (c-d) k-point resolved magneto-anisotropy energy (MAE) isosurface at ±0.03 meV/f.u. for bulk $CrPd_5I_2$ with switching the magnetic axis from [001] to [100] with SOC. (e-f) Band structure with SOC for bulk and 1L $CrPd_5I_2$ with the highest valence band (N) shown in red. (g-h) Weyl nodal lines of bulk and 1L $CrPd_5I_2$ between band N and N+1.

# Supplementary Information

# Design van der Waals Layered Quantum Materials of MPd$_5$I$_2$ (M=Ga, In and 3$d$ Transition Metals)

Niraj K. Nepal[1], Tyler J. Slade[1], Joanna M. Blawat[2], Andrew Eaton[3], Johanna C. Palmstrom[2], Benjamin G. Ueland[1,3], Adam Kaminski[1,3], Robert J. McQueeney[1,3], Ross D. McDonald[2], Paul C. Canfield[1,3,#] and Lin-Lin Wang[1,3,*]

[1]Ames National Laboratory, U.S. Department of Energy, Ames, IA 50011, USA

[2]National High Magnetic Field Laboratory, Los Alamos National Laboratory, Los Alamos, NM 87545, USA

[3]Department of Physics and Astronomy, Iowa State University, Ames, IA 50011, USA

#canfield@ameslab.gov

*llw@ameslab.gov

Supplementary Figure 1: Phase stability and structural energies of MPd$_5$I$_2$ calculated in PBEsol

Supplementary Figure 2: Phonon band dispersion of bulk and 1L MPd$_5$I$_2$ (M=Ti, In and Cr) calculated in opt86b

Supplementary Figure 3: Electronic band structures of other MPd$_5$I$_2$ calculated in opt86b with spin-orbit coupling (SOC)

Supplementary Figure 4: Electronic band structures of bulk and 1L TiPd$_5$I$_2$ calculated with spin-orbit coupling (SOC) using different exchange-correlation functionals

Supplementary Figure 5: Electronic band structures of bulk and 1L InPd$_5$I$_2$ calculated with spin-orbit coupling (SOC) using different exchange-correlation functionals

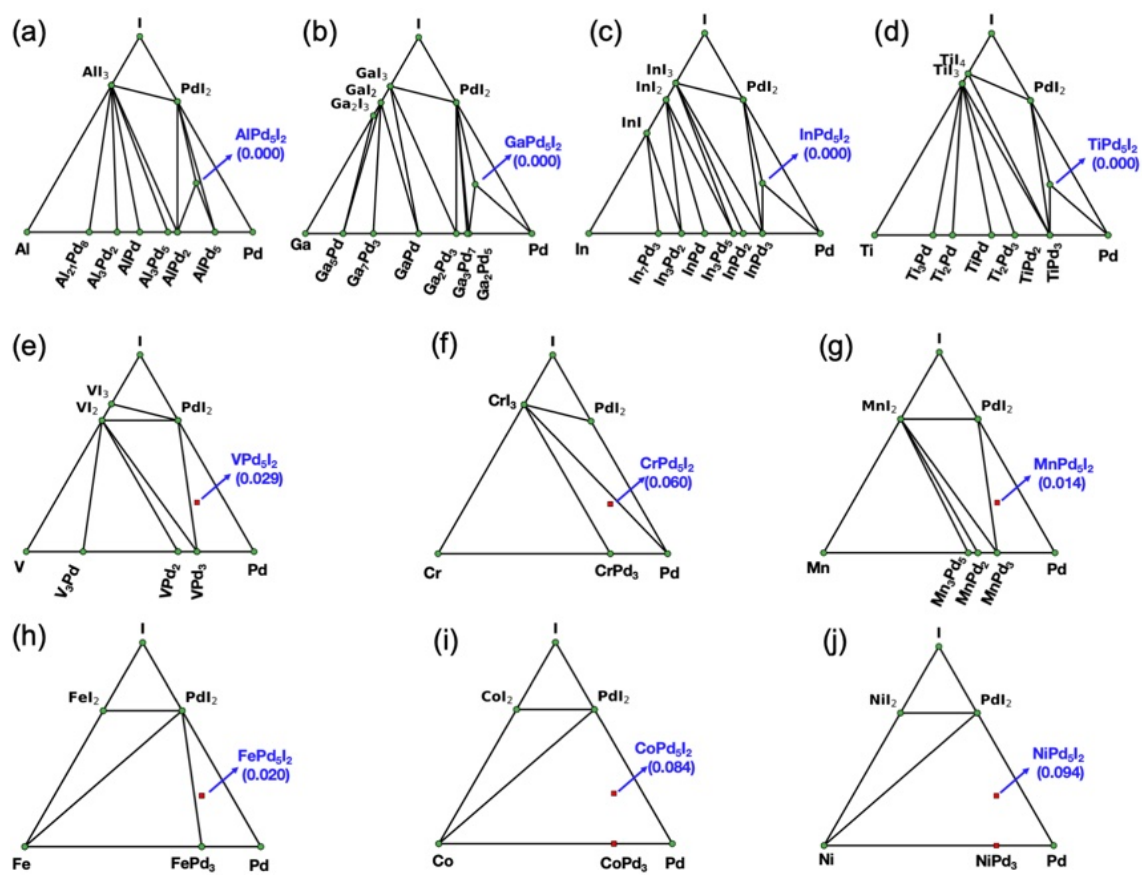

Supplementary Figure 1. Phase stability and structural energies of MPd$_5$I$_2$ calculated in PBEsol. (a)-(j) are for MPd$_5$I$_2$ (M=Al, Ga, In, Ti, V, Cr, Mn, Fe, Co and Ni). The compounds on the ground state hull are labeled as green dots. The MPd$_5$I$_2$ are indicated with arrows and their respective hull energies are listed in parenthesis.

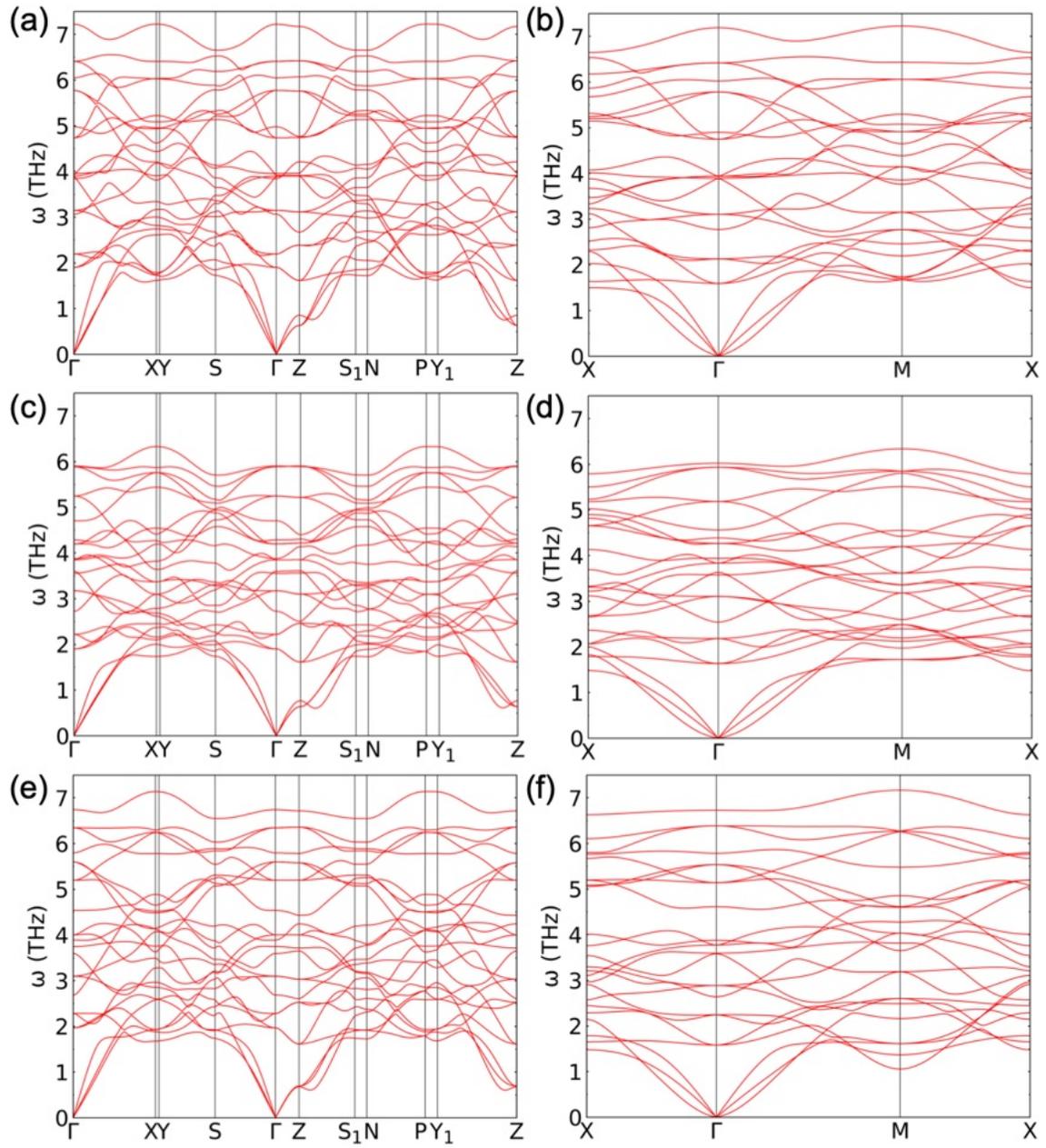

Supplementary Figure 2. Phonon band dispersion of bulk and 1L $MPd_5I_2$ (M=Ti, In and Cr) calculated in opt86b. (a-b) are for bulk and 1L $TiPd_5I_2$. (c-d) are for bulk and 1L $InPd_5I_2$. (e-f) are for bulk and 1L ferromagnetic $CrPd_5I_2$.

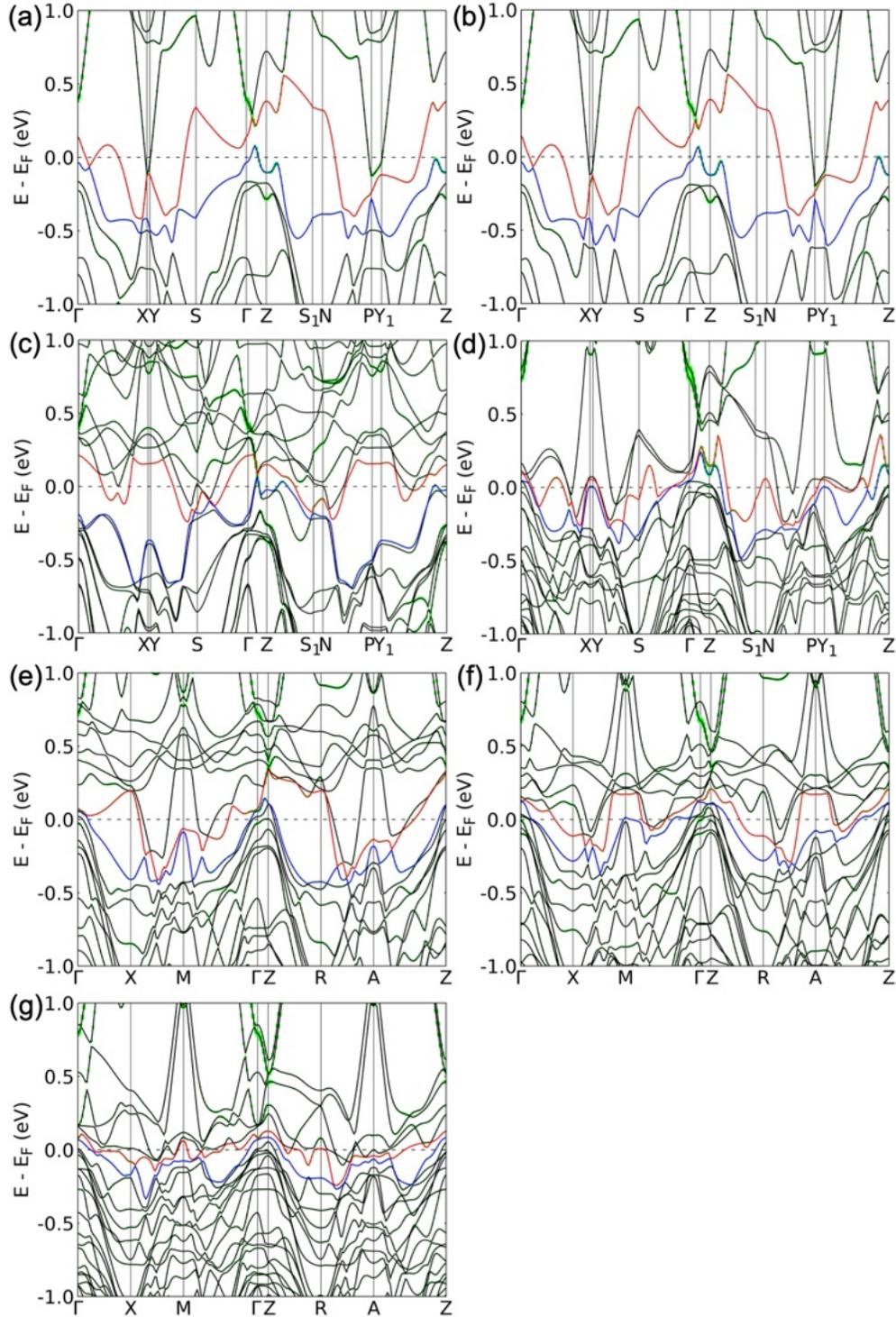

Supplementary Figure 3. Electronic band structures of other $MPd_5I_2$ calculated in opt86b with spin-orbit coupling (SOC). The highest valence band (N) and band below (N-2) shown in red and blue, respectively. Green shade stands for I $p_z$ orbital projection. (a-g) are for $MPd_5I_2$ with M=Al, Ga, V, Mn, Fe, Co and Ni, where Al and Ga in (a-b) are non-magnetic, V and Mn in (c-d) are ferromagnetic, and Fe, Co and Ni in (e-g) are anti-ferromagnetic.

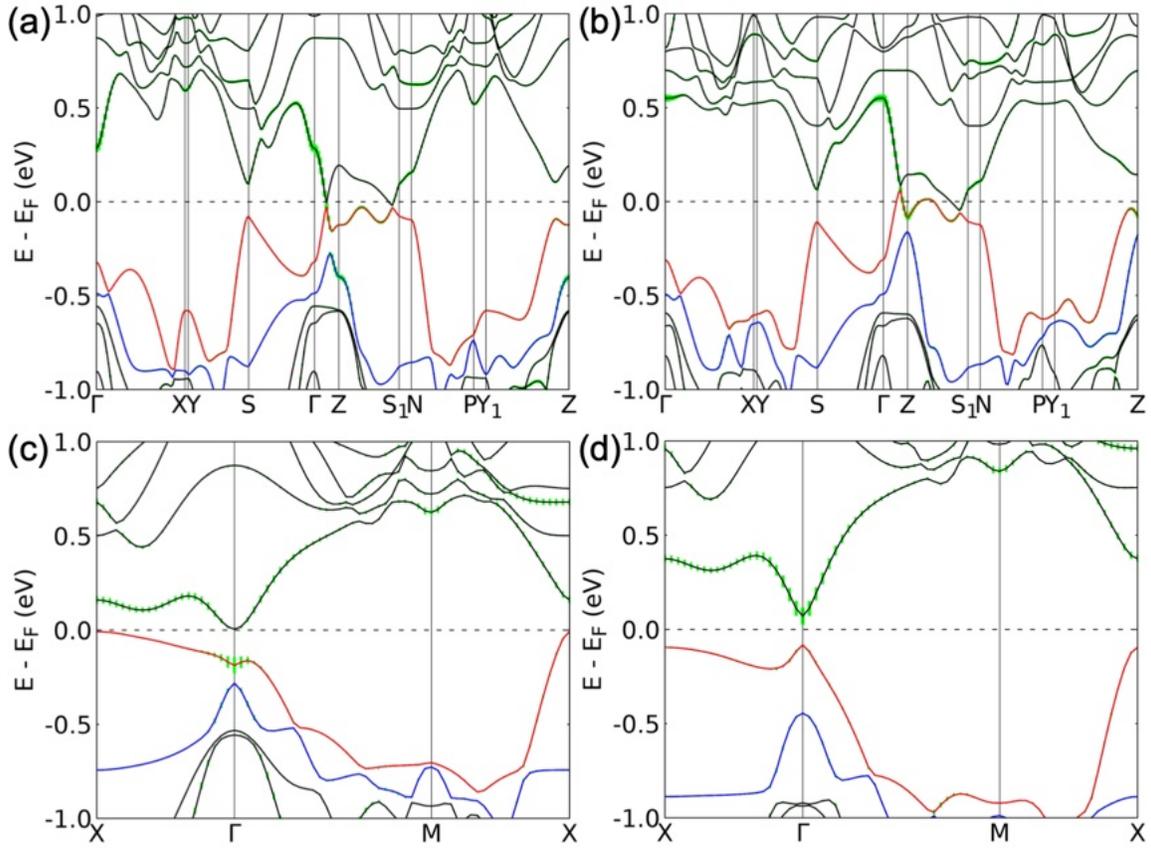

Supplementary Figure 4. Electronic band structures of bulk and 1L TiPd$_5$I$_2$ calculated with spin-orbit coupling (SOC) using different exchange-correlation functionals. (a) Bulk in r2SCAN+rVV10, (b) bulk in mBJ, (c) 1L in r2SCAN+rVV10 and (d) 1L in HSE06 exchange-correlation functionals. The highest valence band (N) and band below (N-2) are shown in red and blue, respectively. Green shade stands for I $p_z$ orbital projection.

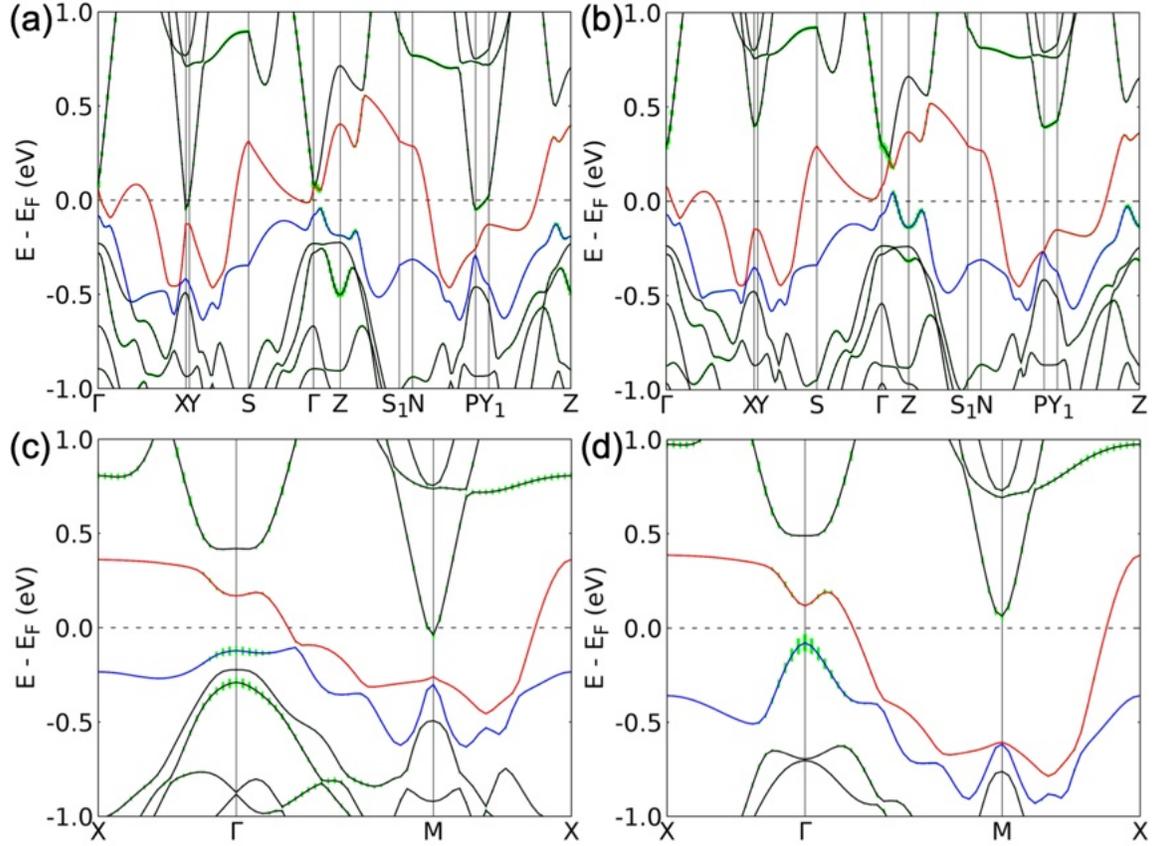

Supplementary Figure 5. Electronic band structures of bulk and 1L InPd$_5$I$_2$ calculated with spin-orbit coupling (SOC) using different exchange-correlation functionals. (a) Bulk in r2SCAN+rVV10, (b) bulk in mBJ, (c) 1L in r2SCAN+rVV10 and (d) 1L in HSE06 exchange-correlation functionals. The highest valence band (N) and band below (N-2) are shown in red and blue, respectively. Green shade stands for I $p_z$ orbital projection.

31